\documentclass[twocolumn]{aastex631}

\usepackage{comment}
\usepackage{booktabs} 
\usepackage{longtable}  

\begin{document}

\title{SN Progenitors of SNRs in OB Associations in the LMC. I. Assessment of Methodology Using N63A
}

%\correspondingauthor{August Muench}
%\email{greg.schwarz@aas.org, gus.muench@aas.org}

\author[0000-0003-2887-6381]{Yuka Terada}
\affil{Department of Physics, National Sun Yat-Sen University, No. 70, Lien-Hai Road, Kaohsiung City 80424, Taiwan, R.O.C.}

\author[0000-0003-3667-574X]{You-Hua Chu}
\affil{Department of Physics, National Sun Yat-Sen University, No. 70, Lien-Hai Road, Kaohsiung City 80424, Taiwan, R.O.C.}
\affil{Institute of Astronomy and Astrophysics, Academia Sinica, Taipei 10617, Taiwan, R.O.C.}

\begin{abstract}
Color--magnitude diagrams (CMDs) can constrain the progenitor masses of supernova remnants (SNRs), but mass estimates for massive stars are affected by limited temperature sensitivity, extinction, crowding, and nebular contamination.
We use the Large Magellanic Cloud (LMC) SNR N63A and its OB association LH83 as a benchmark to evaluate CMD-based mass estimates.
Among the tested diagnostics, the extinction-corrected MCPS $B$ vs.\ $U-B$ CMD provides the most useful constraint on the massive-star population.
The $V$ vs.\ $B-V$ CMD is less reliable, while the reddening-free $W$--$Q$ diagram does not improve individual mass estimates because of its larger photometric uncertainties.
Gaia EDR3 recovers bright stars missing or saturated in MCPS but provides weaker separation among the hottest stars.
For seven spectroscopically classified stars, the $B$ vs.\ $U-B$ CMD-based masses are lower than the HR-diagram masses inferred using the same MIST tracks by $8.0~M_\odot$ on average, with a scatter of $7.0~M_\odot$.
Isochrone comparisons suggest that most of the massive
stars in LH83 formed about 7--15 Myr ago, complicating progenitor mass estimates based on the present-day mass function.
Nevertheless, the masses of D243-1a and D243-2, stellar lifetimes predicted by the MIST models, and the empirical bubble-size--progenitor-mass relation suggest a progenitor mass of order 25--30~$M_\odot$.
Our results highlight the potential and limitations of this
approach for studying other LMC SNRs in OB associations.

\end{abstract}



\section{Introduction} \label{sec:intro}

Broadly speaking, there are two categories of supernovae (SNe): the core-collapse (CC) SNe originating from massive stars and the Type Ia SNe from white dwarfs in close binary systems.  
The CC and Type Ia SNe can be distinguished from the presence or absence of lines of  hydrogen, helium, and metals in their spectra  \citep{Doggett1985}.
When a SN has faded away and a young supernova remnant (SNR) is formed, the SN type can  still be diagnosed from the metal abundance revealed by X-ray spectra of the SN ejecta within the SNR \citep{Hughes1995}.  
In evolved SNRs, where SN ejecta has cooled below X-ray-emitting temperatures, the types of their SNe have been assessed by the stellar and interstellar environments, as CC SNe are more likely to be near blue stars and \ion{H}{2} regions \citep{Chu1988} or associated with recent star formation \citep{Badenes2009}.

A number of SNRs in the Large Magellanic Cloud (LMC) are projected within OB associations, and the SN progenitors were probable members of the OB associations \citep{Chu1997}.  
If a host OB association resulted from a coeval starburst, its SN progenitors that have already exploded would be expected to have had initial masses greater than those of the present-day  main-sequence members of the same population.
Based on this rationale, it has been estimated that the SN progenitor of the SNR N63A was at least  30 $M_\odot$ \citep{vandenBergh1980}.  

The availability of surveys of stars in the LMC, such as the Magellanic Cloud Photometric Survey \citep[MCPS;][]{Zaritsky2004} and the Gaia mission \citep{Gaia2020}, makes it easy to construct color-magnitude diagrams (CMDs) and analyze the stellar population.  
Through comparisons with stellar evolutionary tracks and isochrones, it is possible to estimate the main-sequence turn-off and present-day mass function.  This technique has been used to assess the SN progenitor mass of SNR B0532$-$67.5 \citep{Li2022}, stellar energy feedback in superbubble N51D \citep{Cooper2004} and supergiant shell LMC 1
\citep{Li2025}.

\begin{figure*}[htbp]
    \centering
    \includegraphics[width=\textwidth]{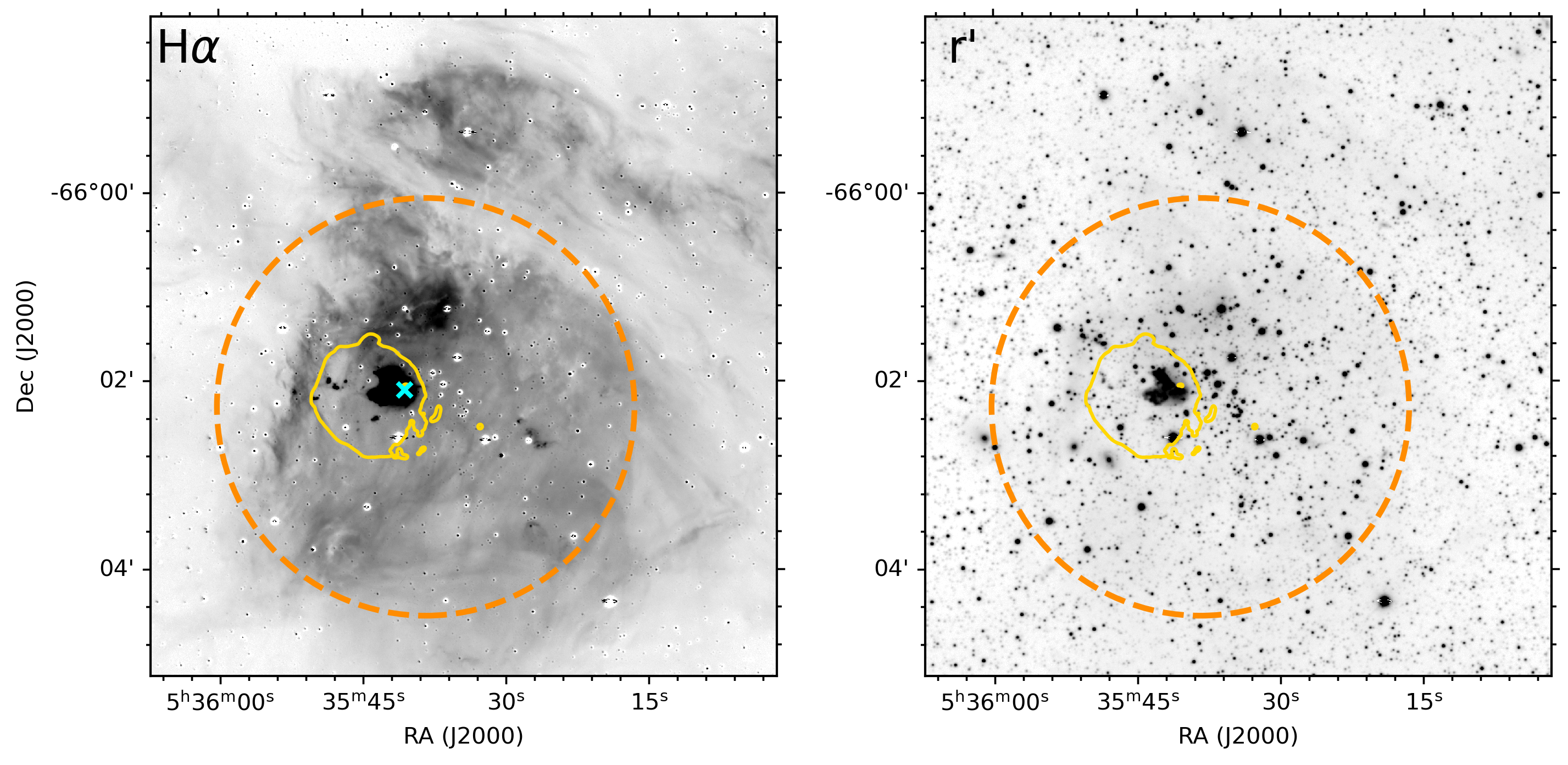}
    \caption{Left: DeMCELS continuum-subtracted H$\alpha$ image showing the \ion{H}{2} region N63. Right: DeMCELS red continuum image showing the stars. The yellow contour marks the outer boundary of the diffuse X-ray emission detected by Chandra observation of the SNR N63A. The dashed orange circle marks the extent of the OB association LH83 defined by \citet{Lucke1970}. The cyan cross marks the cataloged position of HD~271389.
    }
    \label{fig:n63a_ha_continuum}
\end{figure*}

Very massive main-sequence stars have high effective temperatures and their radiation peaks in far ultraviolet; thus, using CMDs constructed in optical wavelengths to examine the most massive stars may be subject to uncertainties.
Such uncertainties must be assessed quantitatively and critically before applying the photometric CMD analysis to assess progenitor masses of supernovae in OB associations. 
For this purpose, we choose to use the SNR N63A as a benchmark to assess  the effectiveness of the CMD method.  
See Figure~\ref{fig:n63a_ha_continuum} for the location of the SNR N63A relative to the OB association LH83 \citep{Lucke1970} and its photoionized \ion{H}{2} region N63.

N63A is a well-studied SNR, with imaging and spectroscopic observations available at multiple
wavelengths \citep[e.g.,][]{Levenson1995, Warren2003, Williams2006}. 
More importantly, the OB association LH83 in N63 has been extensively characterized both photometrically
and spectroscopically by \citet{Oey1996spec}. 
The CMD-based mass estimates can therefore be compared with the 
spectroscopy-based estimates reported by \citet{Oey1996spec} to evaluate the uncertainties of the CMD method. 
Based on this benchmark test, we then apply the same approach to four other SNRs in OB associations in the LMC in an accompanying paper.

This paper is organized as follows. 
Section~\ref{sec:data} describes the photometric and imaging datasets used in our analysis, including the MCPS and Gaia catalogs, as well as the methodology for assessing stellar masses in the CMDs.
Section~\ref{sec:results} presents our detailed investigation of the stellar population
associated with N63A, comparing the  results from multiple CMDs and evolutionary tracks with 
previous spectroscopic studies,  in order to evaluate the reliability of our methodology.
Section~\ref{sec:discussion} discusses the age of LH83, the implications for the N63A progenitor mass, and the limitations of the CMD-based method.
Finally, Section~\ref{sec:conclusion} summarizes our main conclusions.

\section{Data and Methodology}\label{sec:data}

To investigate the stellar population associated with N63A, we have used datasets of both stars and the interstellar medium (ISM).  
The stellar data are the main focus of  this section and will be described in detail.  The multiwavelength imaging data of  the ISM are only used to illustrate the physical structures of the ISM and their relative locations to the stars.  
These include H$\alpha$ and red continuum images from the Dark Energy Camera Magellanic Clouds Emission-Line Survey \citep[DeMCELS;][]{Points2024} to show the distribution of gas and stars, and archival Chandra X-ray observations to mark the boundary of the SNR.
We refer to \citet{Warren2003} and \citet{Karagoz2023} for detailed X-ray studies of the SNR N63A. 

Below we describe the datasets and analysis methods used to investigate the stellar  population associated with N63A, including the construction of CMDs, extinction  corrections, and comparisons with stellar evolutionary models used to estimate stellar masses and ages.

\subsection{Stellar Data}
\label{subsec:photo_data}

We use photometric data from the MCPS \citep{Zaritsky2004} and Gaia Early Data Release 3 \citep[EDR3;][]{Gaia2016, Gaia2020}  for the stellar population analysis. 
For MCPS, we use the $U$, $B$, and $V$ magnitudes.
For Gaia, we use the $G$, $G_{\rm BP}$, and $G_{\rm RP}$ magnitudes. 

Before using the MCPS photometry, it is important to note two issues cautioned by \citet{Zaritsky2004}.
First, the photometry of bright stars ($B<13.5$ or $V<13.5$) may suffer from substantial uncertainties due to imperfect PSF fitting. 
Although MCPS incorporates photometry from \citet{Massey2002} for many bright stars, some objects do not have 
such replacements available and may therefore retain unreliable measurements. 
The treatment of these stars in our analysis is described in Section~\ref{subsec:CMD_MCPS}.

Second, the MCPS $U$-band photometry deviates from that of \citet{Massey2002} and requires a color-dependent correction in order to conform with the latter: $\Delta U = -0.5(B-V-0.7)^2 + 0.23$ for stars with $(B-V)<1$, and $\Delta U = 0.05$ otherwise. 
We applied this correction to the MCPS $U$ magnitudes of blue stars, except for bright stars whose catalog flags indicate that corrected photometry had already been adopted in the MCPS catalog.

Before comparing CMD-based mass estimates among different datasets, we also examine the consistency between the MCPS photometry and the published $UBV$ photometry of \citet{Oey1996UBV}. 
The $V$-band magnitudes from the two catalogs were crossmatched using a positional tolerance of $2^{\prime\prime}$.
This matching radius was adopted to recover the brightest massive stars, which were frequently missed when a smaller matching radius was used because of small positional offsets between the catalogs.

\begin{figure}[htbp]
  \centering
  \includegraphics[width=0.95\linewidth]{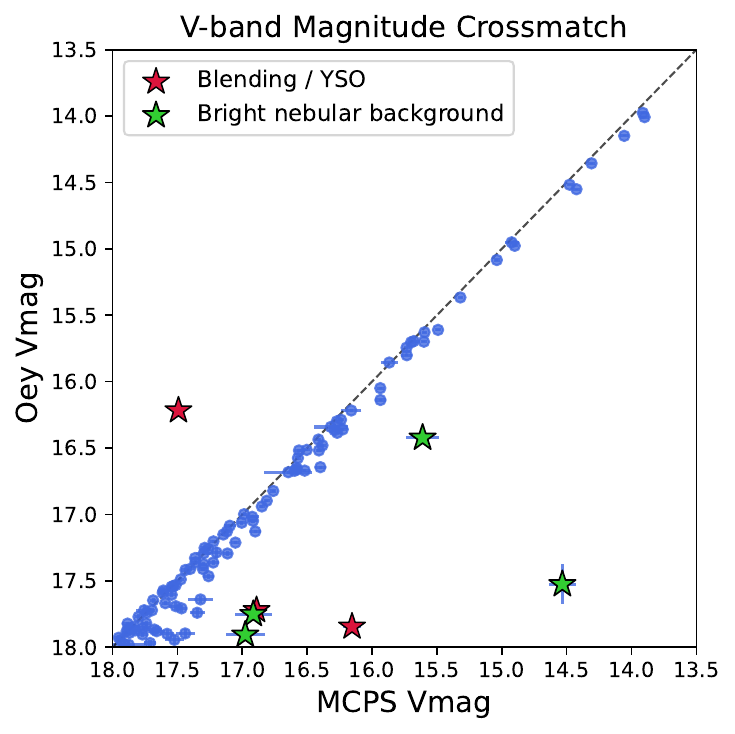}
\caption{
  Comparison of $V$-band magnitudes between the photometric catalogs of \citet{Oey1996UBV} and the MCPS \citep{Zaritsky2004} for stars in the LH83 region brighter than $V=18$ mag. 
  The dashed line indicates the equality line ($V_{\rm MCPS}=V_{\rm Oey}$). Filled blue circles show matched sources. 
  Horizontal and vertical error bars represent the MCPS and Oey photometric uncertainties, respectively, although most are smaller than the plotted symbols. 
  Red stars indicate sources with large magnitude differences ($|\Delta V| > 0.46$ mag), and green stars mark sources in regions of bright nebulosity; see text for details.
  }
  \label{fig:LH83_vmag_comparison}
\end{figure}

Figure~\ref{fig:LH83_vmag_comparison} compares the $V$-band photometry of stars brighter than $V=18$ mag in the Oey and MCPS catalogs, excluding sources affected by the known bright-star photometric issue discussed above.
Most sources are distributed close to the equality line, indicating that the photometric measurements in the two catalogs are generally consistent. 
Although a matching radius of $2^{\prime\prime}$ may introduce ambiguous associations for some faint sources, the brightest stars of primary interest in this study show good agreement between the two datasets.

Several outliers are present. 
To investigate the cause of these discrepancies, we examined the higher-resolution DeMCELS red-continuum images of ten sources with the largest magnitude differences.
Four of these sources are located within the bright nebula around HD~271389, whose cataloged position is marked in Figure~\ref{fig:n63a_ha_continuum}, and the MCPS photometric measurements are severely compromised by the very high nebular background.
Five sources have close neighboring stars, and the blending with them most likely  affects the photometry, as some of the sources' coordinates fall between two close neighboring stars. 
The remaining source has been classified as a young stellar object with mid-infrared excess \citep{Caulet2008}.  
Intrinsic photometric variability may be expected for such young objects.

For all matched stars brighter than $V=18$ mag, the mean magnitude difference is $0.15 \pm 0.42$ mag. 
After excluding these identified outliers, the difference decreases to $0.08 \pm 0.10$ mag.
This substantial reduction in the dispersion indicates that the large discrepancies are dominated by a small number of blended or variable sources, and that the MCPS photometry is broadly consistent with the photometry used by \citet{Oey1996UBV}. 
A small residual offset remains, with the MCPS magnitudes being systematically brighter by approximately 0.08 mag on average, which may reflect minor differences in the photometric calibration adopted by the two studies.

While the overall agreement between the \citet{Oey1996UBV} and MCPS catalogs is good, this comparison demonstrates that photometric measurements of individual stars can be unreliable in crowded or very bright nebular regions. 
Because the primary goal of this study is to identify the most massive stars in LH83, candidate massive stars selected from the CMDs are further examined using higher-resolution continuum images to assess possible blending and source confusion.

In addition to the photometric catalogs, spectroscopic observations of LH83 were reported by \citet{Oey1996spec}. 
In that work, stellar effective temperatures and bolometric magnitudes were derived from spectral classifications, and stellar masses were estimated using evolutionary tracks. 
These spectroscopic results provide an important reference for evaluating the CMD-based mass estimates derived in this study.

To make the reference-sample matching reproducible, Table~\ref{tab:oey_crossid} lists the cross-identifications of the spectroscopically classified Oey stars with the MCPS and Gaia EDR3 catalogs.
The table gives the reference-star coordinates, nearest-catalog matching separations, Gaia EDR3 source identifiers, and flags for potentially ambiguous or crowded cases.
All eight reference stars have MCPS and Gaia counterparts within $2^{\prime\prime}$.
However, D243-1a and D243-1b share the same catalog counterpart in the present matching, and D243-5 has multiple Gaia EDR3 sources within $2^{\prime\prime}$.
These cases are flagged in the table.

\begin{deluxetable*}{lcccccccl}
\tablecaption{
Cross-identification of the spectroscopic reference stars from \citet{Oey1996spec}.
}
\label{tab:oey_crossid}
\tablewidth{\textwidth}
\tablehead{
\colhead{ID} &
\colhead{R.A.} &
\colhead{Decl.} &
\colhead{\shortstack{MCPS sep.\\($^{\prime\prime}$)}} &
\colhead{\shortstack{Gaia EDR3\\source ID}} &
\colhead{\shortstack{Gaia sep.\\($^{\prime\prime}$)}} &
\colhead{\shortstack{$N_{\rm Gaia}$\\within $2^{\prime\prime}$}} &
\colhead{\shortstack{Match\\flag}} &
\colhead{\shortstack{Photometry\\used}}
}
\startdata
D243-1a & 83.921958 & -66.043583 & 1.70 & 4660677546464879872 & 0.66 & 1 & shared counterpart & adopted $UBV$ \\
D243-1b & 83.921958 & -66.043583 & 1.70 & 4660677546464879872 & 0.66 & 1 & shared counterpart & -- \\
D243-2  & 83.884417 & -66.043889 & 0.81 & 4660677580824616576 & 0.66 & 1 & clean & adopted $UBV$ \\
D243-3  & 83.896708 & -66.029417 & 1.23 & 4660677718263585024 & 0.99 & 1 & clean & adopted $UBV$ \\
D243-5  & 83.901125 & -66.020778 & 1.12 & 4660677718254644096 & 0.31 & 5 & crowded & adopted $UBV$ \\
D243-6  & 83.902833 & -66.034139 & 1.13 & 4660677615184370304 & 0.85 & 1 & clean & MCPS \\
D243-9  & 83.865542 & -66.043972 & 0.85 & 4660677580827123328 & 0.74 & 1 & clean & MCPS \\
D243-12 & 83.975750 & -66.058389 & 1.12 & 4660676790550624512 & 0.88 & 1 & clean & MCPS \\
\enddata
\tablecomments{
Coordinates are those of the spectroscopic reference stars from \citet{Oey1996spec}.
The MCPS and Gaia separations are angular offsets between the reference position and the nearest catalog counterpart.
$N_{\rm Gaia}$ gives the number of Gaia EDR3 sources within $2^{\prime\prime}$ of the reference position.
All listed stars have one MCPS counterpart within $2^{\prime\prime}$.
The match flag indicates potentially ambiguous or crowded catalog associations.
D243-1a and D243-1b share the same MCPS and Gaia EDR3 catalog counterparts in the present matching and cannot be treated as independent photometric sources. Therefore, no separate CMD-based mass is assigned to D243-1b. D243-5 is flagged as crowded because multiple Gaia EDR3 sources are found within $2^{\prime\prime}$.
``Adopted $UBV$'' indicates that the bright-star photometry used for the CMD analysis was taken from the literature-based adopted values described in Section~\ref{subsec:CMD_MCPS}, rather than from the original MCPS measurement.
}
\end{deluxetable*}

The information used in this paper is not homogeneous across all parts of the analysis.
The broad stellar population analysis and the construction
of the MCPS CMDs rely primarily on MCPS photometry, after applying the $U$-band correction described above.
However, the benchmark comparison of the CMD-based mass estimates uses external information: the spectroscopic classifications, $\log T_{\rm eff}$, and $M_{\rm bol}$ values from \citet{Oey1996spec}, and in some bright-star cases the historical $UBV$ photometry adopted instead of the original MCPS measurements.
Therefore, when the same CMD-based method is applied to other SNRs without comparable spectroscopic or historical photometric data, the analysis can still identify candidate massive populations from MCPS-like photometry, but it cannot provide the same object-by-object benchmark test against spectroscopic stellar parameters.

\subsection{Photometric Analysis} \label{sec:analysis}

The stellar population analysis of LH83 is affected by spatially variable extinction associated with the nebular environment around N63A. 
We therefore analyze both extinction-corrected CMDs and reddening-free CMDs in order to evaluate the effects of differential extinction on the stellar mass estimates.

\subsubsection{CMD Construction}

The target region for the photometric analysis is the OB association LH83 defined by \citet{Lucke1970}. 
The extent of the association can be approximated by a circular region centered at (RA, Dec) = (83.91$^\circ$, $-66.038^\circ$) with a radius of 0.037$^\circ$, as shown in Figure~\ref{fig:n63a_ha_continuum}. 
To examine the surrounding stellar population, we also define a larger annular background region with radii from 0.037$^\circ$ to 0.074$^\circ$.
Photometric data of stars from the target region and the background region are retrieved from
MCPS and Gaia EDR3.

Because individual stellar extinctions are not available for all stars in the field, we adopt $A_V = 0.35$ mag as the fiducial extinction value for the main CMD analysis.
This value corresponds to the average extinction derived for hot stars in LH83 by \citet{Zaritsky2004}.

We adopt the LMC extinction coefficients from \citet{Wang2023}, who derived an average optical--infrared interstellar extinction law for the Magellanic Clouds using red supergiants and classical Cepheids as extinction tracers.
The extinction law itself describes the wavelength dependence of interstellar dust extinction and is not a property of the target stellar spectral type.
However, broad-band extinction coefficients can depend weakly on the assumed stellar SED through the effective wavelength of each filter.
This possible SED dependence is most relevant for broad passbands, especially Gaia $G$, $G_{\rm BP}$, and $G_{\rm RP}$.
Therefore, we treat the Gaia CMD as a complementary diagnostic and rely primarily on the MCPS $B$ versus $U-B$ CMD when assessing the upper main sequence and benchmark-star masses.
The MCPS $U$ and $B$ magnitudes were dereddened using $A_U/A_V = 1.492 \pm 0.046$ and $A_B/A_V = 1.282 \pm 0.016$, respectively, while the $V$ magnitudes were corrected using the adopted $A_V$ value.
The Gaia $G_{\rm BP}$, $G_{\rm RP}$, and $G$ magnitudes were dereddened using $A_{G_{\rm BP}}/A_V = 0.985 \pm 0.009$, $A_{G_{\rm RP}}/A_V = 0.605 \pm 0.005$, and $A_G/A_V = 0.731 \pm 0.007$, respectively.
We adopt a distance modulus of $DM = 18.5$ for the LMC.

To quantify how the adopted extinction affects the benchmark-star mass estimates, we also repeated the $B$ versus $U-B$ CMD classification for $A_V = 0.17$, 0.35, 0.50, and 0.65 mag.
These values are not intended to define separate preferred extinction solutions.
Instead, they bracket a plausible range from the lower nebular-extinction estimate to higher plausible stellar extinctions in the LH83/N63A region, with $A_V=0.35$ mag used as the fiducial value in the main analysis.
Table~\ref{tab:av_sensitivity} summarizes the resulting mass bins for the spectroscopic reference stars.
The test shows that some benchmark stars move to higher mass bins when larger stellar extinctions are adopted.
For example, D243-1a and D243-2 move from the 25--40 $M_\odot$ bin at the fiducial $A_V=0.35$ mag to the 40--60 $M_\odot$ bin at $A_V=0.65$ mag, while D243-6 moves from the 15--25 $M_\odot$ bin to the 25--40 $M_\odot$ bin when $A_V$ is increased to 0.50--0.65 mag.
Other stars, such as D243-3, remain in the same mass bin over the tested extinction range.
Thus, extinction variations can affect the comparison with the spectroscopic reference sample, and individual CMD-based masses should be interpreted as approximate and extinction-dependent.

\begin{deluxetable}{lcccc}
\tablecaption{
Sensitivity of the $B$ versus $U-B$ CMD mass bins to the adopted extinction.
}
\label{tab:av_sensitivity}
\tablewidth{0pt}
\tablehead{
\colhead{ID} &
\colhead{$A_V=0.17$} &
\colhead{$A_V=0.35$} &
\colhead{$A_V=0.50$} &
\colhead{$A_V=0.65$}
}
\startdata
D243-1a & 15--25($\sim23$)  & 25--40($\sim28$)  & 25--40($\sim35$)  & 40--60($\sim43$) \\
D243-2  & 15--25($\sim23$)  & 25--40($\sim30$)  & 25--40($\sim38$)  & 40--60($\sim43$) \\
D243-3  & 15--25($\sim17$)  & 15--25($\sim19$)  & 15--25($\sim22$)  & 15--25($\sim24$) \\
D243-5  & 10--15($\sim13$)  & 15--25($\sim16$)  & 15--25($\sim17$)  & 15--25($\sim20$) \\
D243-6  & 15--25($\sim20$)  & 15--25($\sim23$)  & 25--40($\sim28$)  & 25--40($\sim35$) \\
D243-9  & 15--25($\sim19$)  & 15--25($\sim22$)  & 25--40($\sim26$)  & 25--40($\sim37$) \\
D243-12 & 10--15($\sim13$)  & 15--25($\sim16$)  & 15--25($\sim17$)  & 15--25($\sim20$) \\
\enddata
\tablecomments{
Mass bins are given in units of $M_\odot$.
The listed values were obtained by repeating the same $B$ versus $U-B$ CMD classification while varying only the adopted extinction.
The tested $A_V$ values are used to assess the sensitivity of the CMD-based mass estimates to extinction and are not treated as separate preferred extinction solutions.
The fiducial value used in the main CMD analysis is $A_V=0.35$ mag.
D243-1b is not listed because it shares the same catalog photometric counterpart as D243-1a in the present matching, and an independent CMD-based mass bin cannot be assigned from the available catalog photometry.
}
\end{deluxetable}

We construct multiple CMDs using both MCPS and Gaia photometry in order to compare the sensitivity of different filter combinations to the massive stellar population. 
For MCPS, we use the $B$ vs.\ $U-B$ and $V$ vs.\ $B-V$ diagrams. 
For Gaia, we use $G$ vs.\ $G_{\rm BP}-G_{\rm RP}$.

\subsubsection{Reddening-Free CMD}
\label{subsec:CMD_reddening_free}

The CMDs described above are constructed using a single representative extinction value for LH83. 
However, spectroscopic observations by \citet{Oey1996spec} indicate that the extinction varies among individual massive stars within the association. 
To assess the sensitivity of the CMD analysis to differential extinction, we additionally construct a diagram using the parameters $W$ and $Q$:
\begin{equation}
W \equiv V - DM - (B - V)\frac{A_V}{E(B - V)}
\end{equation}

\begin{equation}
Q \equiv (U - B) - (B - V)\frac{E(U - B)}{E(B - V)} ~~,
\end{equation}
where $E(U - B)/E(B - V) = 0.72$ and $A_V/E(B - V) = 3.1$ are adopted.

Because $W$ and $Q$ are constructed from the same $U$, $B$, and $V$ measurements used in the conventional MCPS CMDs, the $W$ versus $Q$ diagram is not an independent observational check. 
Instead, we use it as a complementary diagnostic of how sensitive the inferred massive-star population is to extinction corrections.

The quantities $W$ and $Q$ are reddening-free only under the assumed extinction ratios. 
If the extinction law varies spatially within LH83, or if the adopted ratios are not fully appropriate for the hot-star population, residual extinction effects can remain in the $W$ versus $Q$ diagram. 
Therefore, masses inferred from this diagram are treated as approximate and are used mainly to test whether the same high-mass candidates are recovered when the explicit dependence on a single adopted $A_V$ is reduced.

Because this approximation is most reliable for hot blue stars, we restrict the sample to bright sources with $U < 14.5$ in the MCPS catalog.

The extinction coefficients adopted for the conventional
CMDs imply $A_V/E(B-V) \simeq 3.55$ and
$E(U-B)/E(B-V) \simeq 0.745$.
For the fiducial extinction of $A_V=0.35$ mag,
the differences between these ratios and those adopted
for the $W$--$Q$ diagram result in residual extinction
terms of approximately 0.044 mag in $W$ and
0.002 mag in $Q$.
These residuals are small compared with the propagated
photometric uncertainties in the $W$--$Q$ diagram.

\subsubsection{MIST Evolutionary Tracks and Isochrones}

To estimate stellar masses from the CMDs, we compare the observed stellar distributions with theoretical stellar evolution models from the MESA Isochrones and Stellar Tracks (MIST) project \citep{Dotter2016,Choi2016,Dotter2026,Bauer2026}.
The MIST models are based on stellar evolution calculations performed with the Modules for Experiments in Stellar Astrophysics (MESA) code and provide synthetic photometry for multiple photometric systems, including $UBV$ filters and Gaia filters.

For the evolutionary tracks and isochrones, we used MIST version 1.2 models computed with MESA version 7503.
The adopted model composition is $Z_{\rm init}=0.00451753$, $Y_{\rm init}=0.2557$, $[\mathrm{Fe/H}]=-0.5$, and $[\alpha/\mathrm{Fe}]=0.0$, appropriate for the young stellar population of the LMC.
We used the rotating model set with initial rotation parameter $v/v_{\rm crit}=0.4$.
Synthetic magnitudes were taken from the MIST bolometric-correction tables in the $U$, $B$, and $V$ bands and in the Gaia EDR3 $G$, $G_{\rm BP}$, and $G_{\rm RP}$ bands.

The MIST models are generated without extinction ($A_V = 0$).
Therefore, we apply extinction corrections to the observed photometric data before comparing them with the models.
The extinction-corrected CMDs are then compared with the MIST evolutionary tracks to estimate stellar masses.

For LH83, we use evolutionary tracks with initial masses of 5, 10, 15, 25, 40, and 60~$M_\odot$. These mass values follow the grid adopted by \citet{Oey1996spec}, allowing a direct comparison between the CMD-based mass estimates and the previous spectroscopy-based estimates.
The displayed tracks are the same grid used for assigning CMD-based masses.
No additional evolutionary tracks were generated or numerically interpolated to construct a finer mass grid.
Masses were estimated by visually interpolating between neighboring MIST tracks in the extinction-corrected CMDs.
When a star lay between two tracks, we assigned an approximate mass or mass bin according to its relative position.
Stars lying outside the adopted model locus are reported as such rather than assigned a formal CMD mass.
When different evolutionary phases overlap in a CMD, we did not attempt to distinguish phases using photometry alone; the resulting CMD-based masses are therefore treated as approximate photometric estimates rather than formal best-fit masses.

In the study by \citet{Oey1996spec}, 
spectral classifications were obtained for the brightest stars associated with DEM~L243 (aka N63), and stellar effective temperatures and bolometric magnitudes were estimated from the spectral classifications using published temperature calibrations and bolometric corrections \citep{chlebowski1991, Humphreys1984}. 
The stellar positions in the $\log T_{\rm eff}$--$M_{\rm bol}$ plane were then compared with the evolutionary tracks of \citet{Schaerer1993} to infer initial stellar masses. 

In this work, we use the published positions in the $\log T_{\rm eff}$--$M_{\rm bol}$ plane as an empirical reference and re-evaluate them with the same MIST evolutionary tracks adopted for the CMD analysis.
This provides an HR-diagram MIST mass estimate, $M_{\rm HR,MIST}$, that can be compared directly with the CMD-based MIST estimate, $M_{\rm CMD,MIST}$.

In addition to the evolutionary tracks, we also construct isochrones using the same MIST model set.
The isochrones are used to estimate the characteristic age of the LH83 stellar population.
We use logarithmic ages of $\log(t/{\rm yr}) = 6.48$, 6.70, 6.84, 7.00, 7.18, and 7.30, corresponding to ages of approximately 3--20 Myr.
The isochrones are compared with the MCPS $B$ vs.\ $U-B$ CMD because the upper main-sequence population is most clearly separated in this diagram.

\section{Assessment of CMD-based Mass Estimates in LH83}
\label{sec:results}

In this section, we compare stellar mass estimates derived from different photometric CMDs with the spectroscopic results of \citet{Oey1996spec}. 
We first introduce the spectroscopic reference sample and then examine how well the massive stellar population is recovered in the MCPS, Gaia, and reddening-free CMDs. 
Finally, we discuss the strengths and limitations of each photometric approach.

\subsection{Reference massive-star sample from spectroscopy}
\label{subsec:oey_spec_reference}

To establish a reference sample of massive stars in LH83, we adopt the spectroscopically classified stars 
presented by \citet{Oey1996spec}.
In that study, spectral classifications were obtained for the brightest stars associated with DEM~L243, and stellar effective temperatures and bolometric magnitudes were estimated from the spectral classifications using published temperature calibrations and bolometric corrections. 
The stellar positions in the $\log T_{\rm eff}$--$M_{\rm bol}$ plane were then compared with the evolutionary tracks of \citet{Schaerer1993} to infer initial stellar masses. 
In this work, we use the published $\log T_{\rm eff}$--$M_{\rm bol}$ positions as an empirical reference and compare them with the MIST evolutionary tracks adopted throughout our analysis. 
We refer to the masses reported by \citet{Oey1996spec} as spectroscopy-based evolutionary-track masses.

The stellar properties of the stars used in this comparison are summarized in Table~\ref{tab:oey_cmdphot}. 
We use the stellar identifiers defined by \citet{Oey1996spec} throughout this work. 
Because the photometric CMDs are used to identify the most massive stars in the association, this spectroscopically classified sample provides a benchmark for testing whether the same high-mass population is recovered from the MCPS and Gaia data.

\begin{figure}[htbp]
\centering
\includegraphics[width=0.45\textwidth]{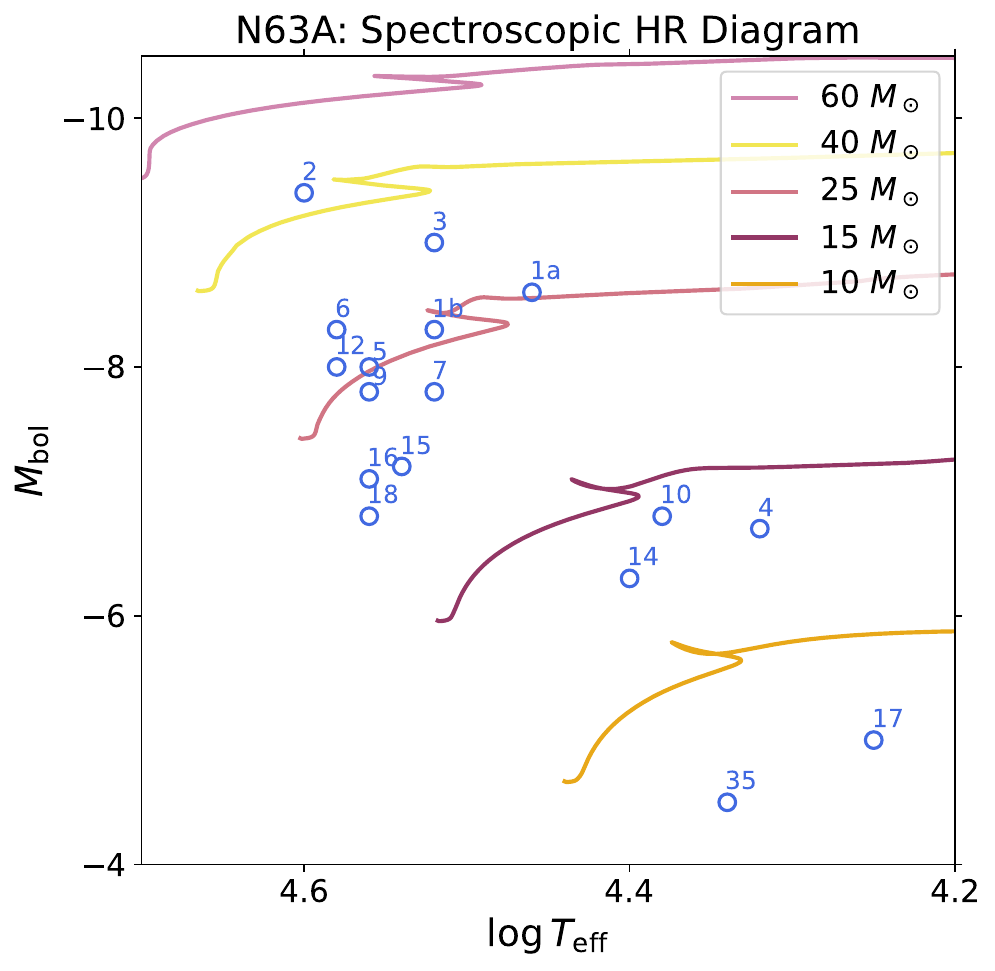}
\caption{
Hertzsprung--Russell diagram of stars in LH83 with spectral classifications from \citet{Oey1996spec}. 
The effective temperatures and bolometric magnitudes were derived from the spectral classifications using published calibrations and bolometric corrections. 
MIST evolutionary tracks for initial masses of 10, 15, 25, 40, and 60~$M_\odot$ are overplotted. 
Blue circles indicate the stars with spectral classifications, and the labels show the source IDs of \citet{Oey1996spec}, with the common prefix D243 omitted for clarity.
}
\label{fig:LH83_oey_specHR}
\end{figure}

Figure~\ref{fig:LH83_oey_specHR} shows the spectroscopically classified stars in the $\log T_{\rm eff}$--$M_{\rm bol}$ plane, together with the MIST evolutionary tracks adopted in this work. 
Most of the spectroscopically classified stars are located between the 15 and 40~$M_\odot$ tracks, while D243-2 lies among the most massive objects in the sample. 
D243-1a is also among the most luminous stars in the sample, with its relatively lower effective temperature being consistent with its evolved supergiant classification.

The masses inferred from the MIST tracks are broadly consistent with the spectroscopy-based evolutionary-track masses reported by \citet{Oey1996spec}. 
See the last two columns of Table~\ref{tab:oey_phot_cmd_mass}.
For the stars in this sample, the HR-diagram masses inferred from the MIST tracks remain close to the values reported by \citet{Oey1996spec} using the \citet{Schaerer1993} tracks.
This comparison provides a limited check of the evolutionary-model dependence.
Therefore, the larger differences between the CMD-based masses and the HR-diagram-based masses are unlikely to be driven primarily by the choice between these two evolutionary grids.
They more likely reflect other uncertainties, including the adopted photometry, extinction corrections, spectroscopic calibrations, and the transformation of the models into the observational CMD plane.

The use of the MIST tracks does not change the identification or relative ranking of the most massive stars in LH83. 
We therefore adopt this spectroscopically classified sample as an empirical benchmark for evaluating the CMD-based mass estimates derived in the following subsections.

\begin{deluxetable*}{cllllcl}
\tablecaption{
Photometric and spectroscopic properties of the massive-star reference sample in LH83, based on the photometry of \citet{Oey1996UBV} and the spectroscopic results of \citet{Oey1996spec}.
}
\label{tab:oey_cmdphot}
\tablehead{
\colhead{No.} & \colhead{ID} & \colhead{$V$} & \colhead{$B-V$} & \colhead{$U-B$} & \colhead{$\log T_{\rm eff}$} & \colhead{Spec Type}
}
\startdata
0 & D243-1a  & 12.822 & $-$0.129 & $-$0.985 & 4.46 & ~~B0 Ib \\
1 & D243-1b  & 13.622 & $-$0.129 & $-$0.985 & 4.52 & ~~O9.5 III \\
2 & D243-2   & 13.067 & $-$0.217 & $-$1.041 & 4.60 & ~~O7 V((f)) \\
3 & D243-3   & 13.370 & $-$0.072 & $-$0.940 & 4.52 & ~~O9.5 III \\
4 & D243-5   & 13.979 & $-$0.296 & $-$0.857 & 4.56 & ~~O7 If + neb \\
5 & D243-6   & 14.008 & $-$0.227 & $-$1.012 & 4.58 & ~~O8 V \\
6 & D243-9   & 14.356 & $-$0.203 & $-$1.014 & 4.56 & ~~O9 V \\
7 & D243-12  & 14.551 & $-$0.124 & $-$0.997 & 4.58 & ~~O8 Ve \\
\enddata
\tablecomments{
Star identifiers, spectral types, and effective temperatures are taken from \citet{Oey1996spec}, while the $UBV$ photometry is taken from \citet{Oey1996UBV}.
The photometric values listed here are the original literature measurements, not necessarily the adopted photometry used for the CMD-based mass estimates.
This sample is used as the spectroscopic reference for comparison with the CMD-based mass estimates in this work.
}
\end{deluxetable*}

\begin{deluxetable*}{llcccccc}
\tablecaption{
Comparison of stellar mass estimates for the massive-star reference sample in LH83.
}
\label{tab:oey_phot_cmd_mass}
\tablewidth{\textwidth}
\tablehead{
\colhead{No.} &
\colhead{ID} &
\colhead{\shortstack{MCPS\\($V$ vs.\ $B-V$)}} &
\colhead{\shortstack{MCPS\\($B$ vs.\ $U-B$)}} &
\colhead{\shortstack{MCPS\\($W$ vs.\ $Q$)}} &
\colhead{\shortstack{Gaia\\($G$ vs.\ $G_{\rm BP}-G_{\rm RP}$)}} &
\colhead{\shortstack{Oey parameters\\+ MIST tracks}} &
\colhead{\shortstack{Oey parameters\\+ Schaerer tracks}}
}
\startdata
0 & D243-1a  & $\sim$25 & $\sim$28 & $\sim$35 & 15--25 & $\sim$25 & $\sim$25 \\
1 & D243-1b  & -- & -- & -- & -- & $\sim$25 & $\sim$25 \\
2 & D243-2   & Left of MS & $\sim$30 & $\sim$23 & $\sim$25 & $\sim$47 & $\sim$50 \\
3 & D243-3   & $\sim$14 & $\sim$19 & $\sim$24 & 10--15 & $\sim$33 & $\sim$35 \\
4 & D243-5   & Left of MS & $\sim$16 & $\sim$13 & -- & $\sim$25 & $\sim$27 \\
5 & D243-6   & $\sim$22 & $\sim$23 & $\sim$24 & 15--25 & $\sim$30 & $\sim$32 \\
6 & D243-9   & 25--40 & $\sim$22 & $\sim$21 & 15--25 & $\sim$23 & $\sim$24 \\
7 & D243-12  & $\sim$10 & $\sim$16 & $\sim$22 & $\sim$8 & $\sim$27 & $\sim$31 \\
\enddata
\tablecomments{
All stellar masses are given in units of $M_\odot$.
The four CMD columns list approximate CMD-based masses or mass bins inferred using the MIST evolutionary tracks.
These values are not formal best-fit masses; they were assigned from the positions of the stars relative to the displayed MIST tracks in each CMD.
The uncertainty of the CMD-based masses is at least comparable to the spacing between adjacent displayed mass tracks and can be larger for stars affected by extinction variations, crowding, or overlapping evolutionary phases.
Entries marked ``Left of MS'' or ``--'' were not assigned formal CMD masses.
The final two columns list masses derived from the published stellar effective temperatures and bolometric magnitudes of \citet{Oey1996spec} using the MIST and \citet{Schaerer1993} tracks, respectively.
}
\end{deluxetable*}

\begin{figure}[htbp]
\centering
\includegraphics[width=0.5\textwidth]{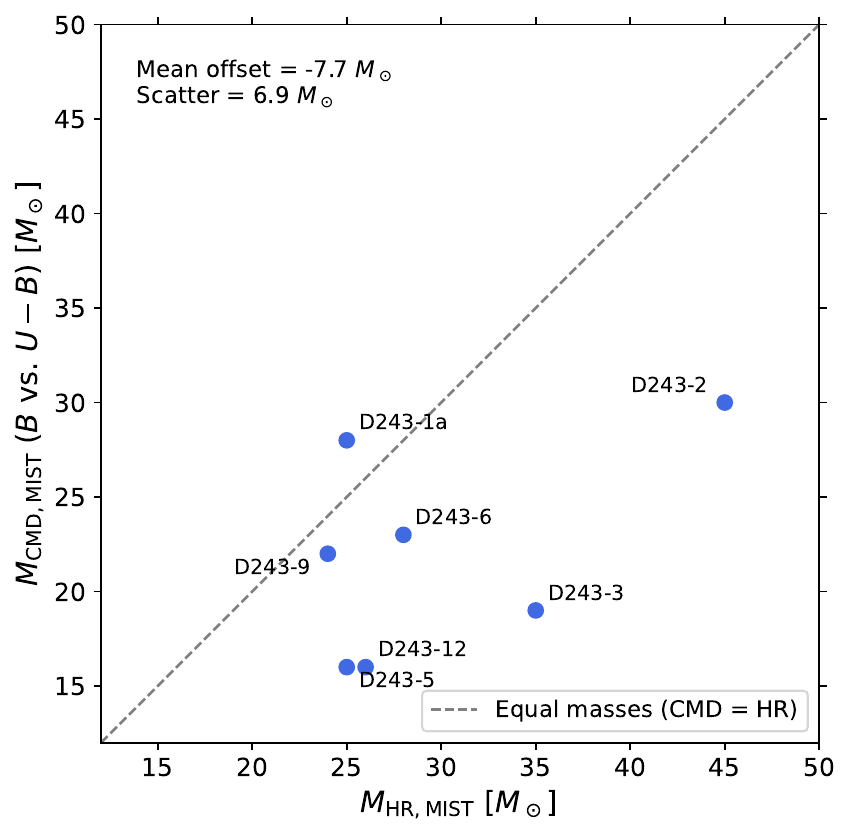}
\caption{
Comparison between the CMD-based masses inferred from the MCPS
$B$ versus $U-B$ diagram and the HR-diagram-based masses inferred
from the published \citet{Oey1996spec} stellar parameters using
the same MIST evolutionary tracks.
The dashed line indicates equal CMD- and HR-diagram-based masses.
For the seven stars, the CMD-based masses are lower than the
HR-diagram-based MIST masses by $8.0~M_\odot$ on average,
with a scatter of $7.0~M_\odot$ and an rms difference of
$10.3~M_\odot$.
Both sets of masses were estimated by visual interpolation
between MIST evolutionary tracks and should be regarded as
approximate rather than formal best-fit estimates.
}
\label{fig:cmd_vs_hr_mass}
\end{figure}

We use the comparison between the CMD-based masses and the HR-diagram masses inferred with the same MIST tracks as the primary quantitative assessment of the photometric method.
This comparison reduces the extent to which differences between evolutionary-track sets can explain the discrepancy.
For the seven stars with usable MCPS $B$ versus $U-B$ CMD estimates, the CMD-based masses are systematically lower than the HR-diagram MIST masses by an average of $8.0~M_\odot$, with a scatter of $7.0~M_\odot$ and an rms difference of $10.3~M_\odot$.
This comparison is shown in Figure~\ref{fig:cmd_vs_hr_mass}.

The HR-diagram masses are not error-free reference values.
They depend on the published spectral classifications, the adopted spectral-type--temperature calibration, bolometric corrections, extinction correction, and uncertainties in the quoted stellar parameters.
Therefore, Figure~\ref{fig:cmd_vs_hr_mass} should be interpreted as an assessment of the consistency and possible systematic bias of the CMD method, rather than as a calibration against exact stellar masses.

\subsection{CMDs from MCPS photometry}
\label{subsec:CMD_MCPS}

\begin{figure}[htbp]
  \centering
  \includegraphics[width=0.4\textwidth]{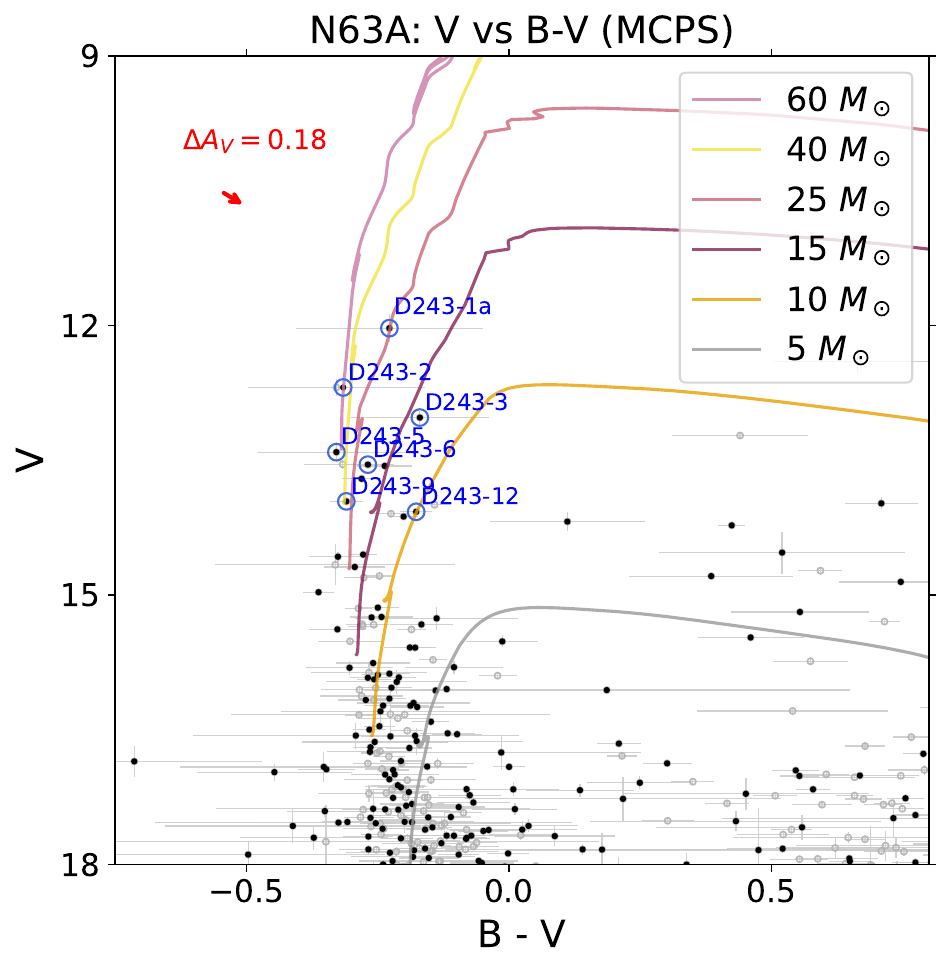}
  \vspace{1em}
  \includegraphics[width=0.4\textwidth]{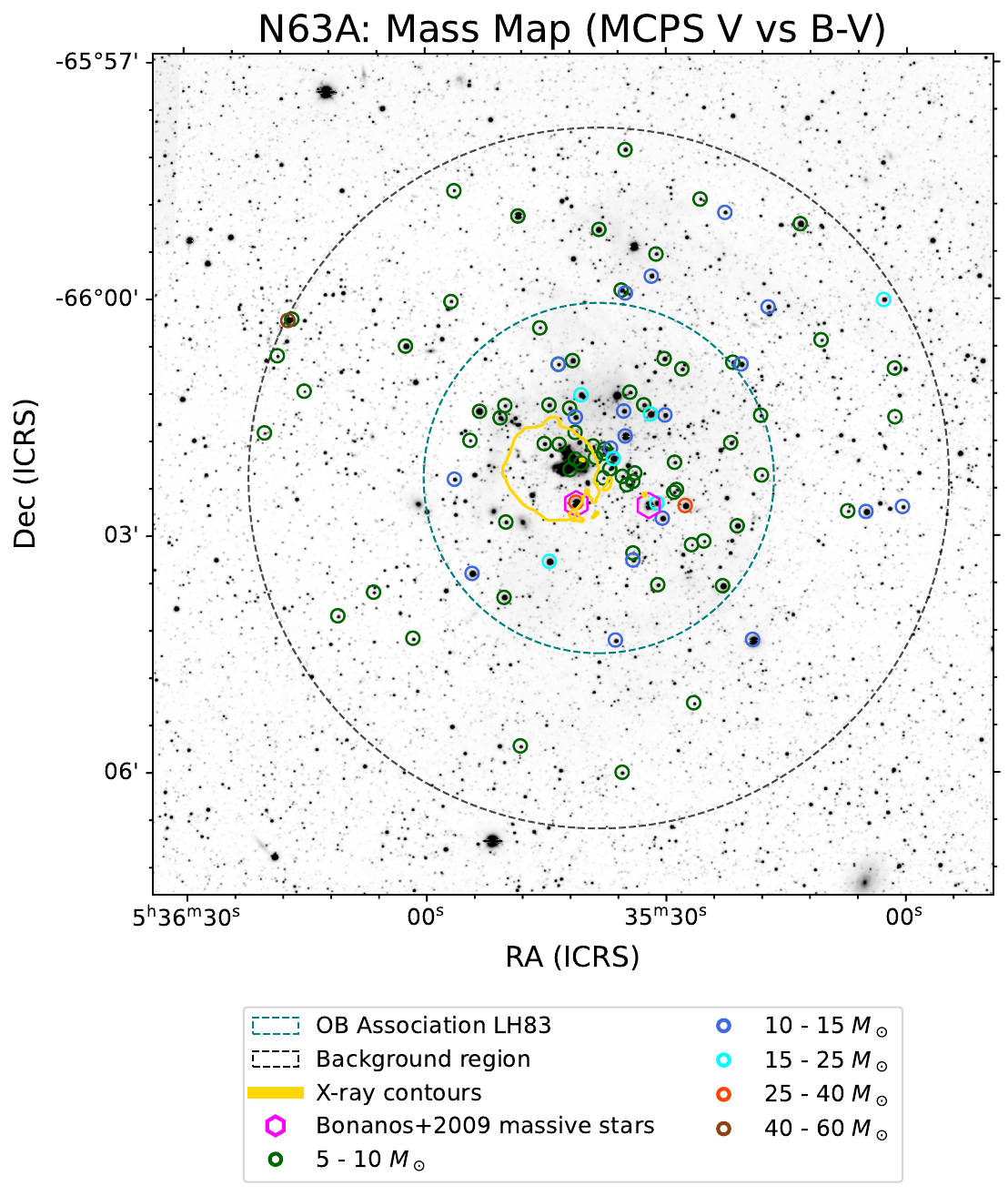}
  \caption{
    LH83: (Top) Color--magnitude diagram ($V$ vs.\ $B-V$) constructed from the MCPS catalog. 
    Stars within the OB association are shown in black, while stars in the surrounding region are shown in gray. 
    MIST evolutionary tracks are overplotted for comparison. 
    (Bottom) Massive star candidates overlaid on the continuum image ($B-V$ case).
    }
  \label{fig:LH83_mcps_bv_summary}
\end{figure}

\begin{figure}[htbp]
  \centering
  \includegraphics[width=0.4\textwidth]{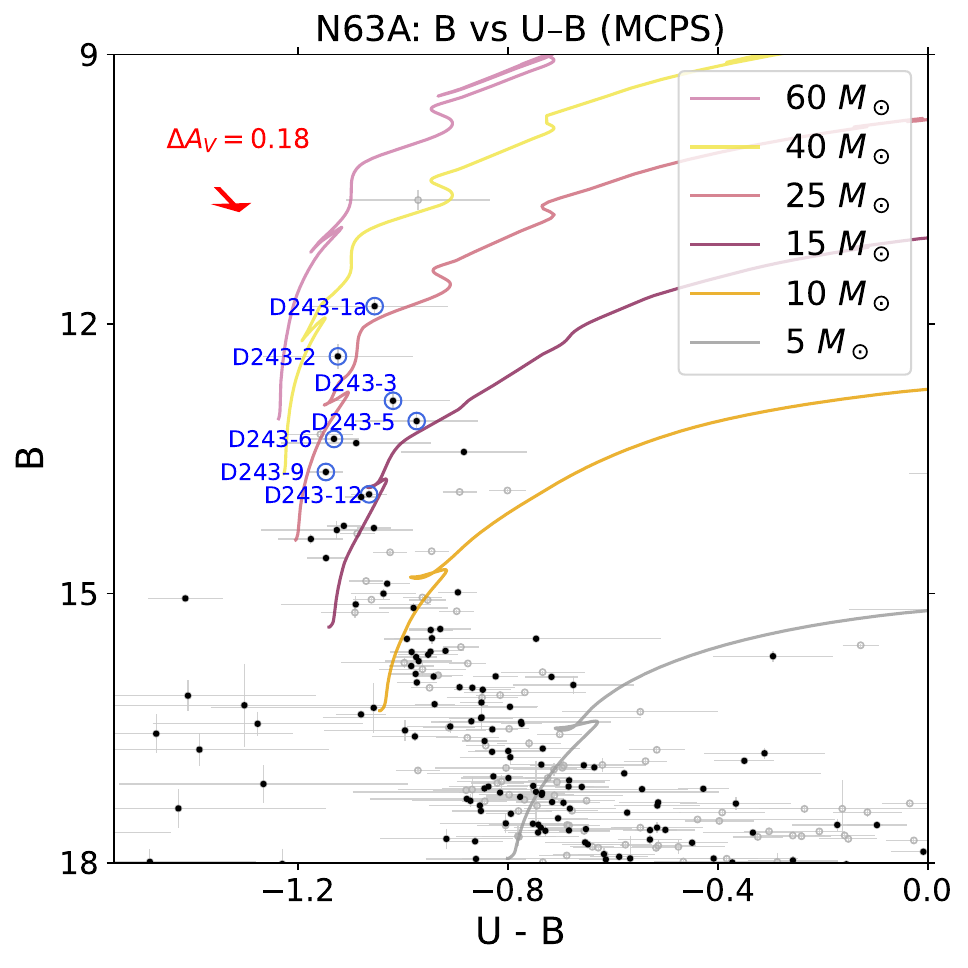}
  \vspace{1em}
  \includegraphics[width=0.4\textwidth]{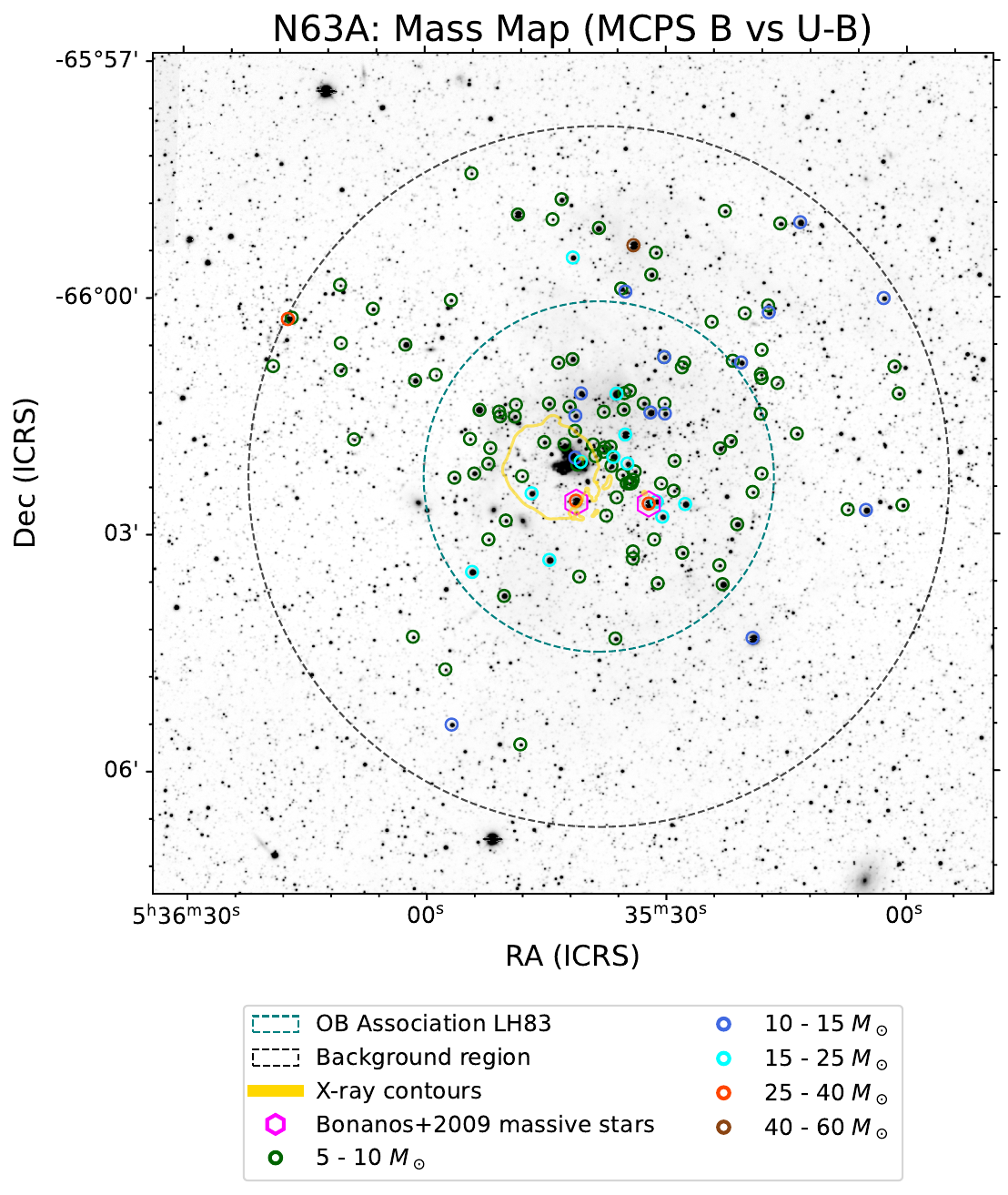}
  \caption{
    LH83: (Top) Color--magnitude diagram ($B$ vs.\ $U-B$) from the MCPS catalog.
    Stars within the OB association are shown in black, while those outside are shown in gray.
    (Bottom) Massive star candidates overlaid on the continuum image.
  }
  \label{fig:LH83_mcps_ub_corr_summary}
\end{figure}

We next examine the MCPS photometry, which provides a deeper sample in the LH83 region than the spectroscopically identified stars alone. 
CMDs constructed from the MCPS catalog are shown in Figures~\ref{fig:LH83_mcps_bv_summary} and \ref{fig:LH83_mcps_ub_corr_summary}. 
As described in Section~\ref{subsec:photo_data}, some bright stars are not covered by the \citet{Massey2002} replacements and retain the original MCPS photometry.
We therefore re-examined these sources before deriving CMD-based
stellar masses.

Table~\ref{tab:bright_star_photometry} summarizes the published photometry considered for the bright LH83 stars whose original MCPS measurements were identified as unreliable. 
In addition to the photometry reported by \citet{Oey1996UBV} and \citet{vandenBergh1980}, we also examined the photometric catalog of \citet{Rousseau1978}, which provides $UBV$ measurements for bright stars in the LMC. 
For D243-1a and D243-2, unambiguous counterparts were identified in all three literature sources. 
The adopted photometry was taken as the median of the available non-MCPS measurements from \citet{Oey1996UBV}, \citet{Rousseau1978}, and \citet{vandenBergh1980}. 
Unfortunately, the source positions in \citet{Rousseau1978} are not sufficiently precise for a reliable identification of the counterparts of D243-3 and D243-5. 
We therefore excluded the \citet{Rousseau1978} measurements for these two stars and adopted the mean of the values reported by \citet{Oey1996UBV} and \citet{vandenBergh1980}. 
The adopted $V$, $B-V$, and $U-B$ values were used to reconstruct the corresponding $U$ and $B$ magnitudes and replace the unreliable MCPS photometry for these four bright stars in the CMD analysis.
These adopted values are distinct from the original Oey photometry listed in Table~\ref{tab:oey_cmdphot}, which is provided for reference.
The MCPS CMD-based mass estimates in Table~\ref{tab:oey_phot_cmd_mass} were obtained using the adopted photometry for these stars.

\begin{deluxetable*}{llccc}
\tablecaption{
Photometric measurements of the brightest stars in LH83.
}
\label{tab:bright_star_photometry}
\tabletypesize{\scriptsize}
\tablewidth{0pt}

\tablehead{
\colhead{Star} &
\colhead{Source} &
\colhead{$V$} &
\colhead{$B-V$} &
\colhead{$U-B$}
}

\startdata
D243-1a & MCPS    & 10.93  & $-0.45$  & 0.71 \\
        & O96     & 12.822 & $-0.129$ & $-0.985$ \\
        & R78     & 12.32  & $-0.12$  & $-0.96$ \\
        & V80     & 12.38  & $-0.15$  & $-0.98$ \\
        & Adopted & $12.38 \pm 0.25$ & $-0.129 \pm 0.015$ & $-0.98 \pm 0.013$ \\
D243-2  & MCPS    & 13.18  & $-1.168$ & $-1.2347$ \\
        & O96     & 13.067 & $-0.217$ & $-1.04$ \\
        & R78     & 12.95  & $-0.17$  & $-1.05$ \\
        & V80     & 13.04  & $-0.23$  & $-1.05$ \\
        & Adopted & $13.04 \pm 0.06$ & $-0.217 \pm 0.03$ & $-1.05 \pm 0.006$ \\
D243-3  & MCPS    & 13.341 & $-0.415$ & $-0.7836$ \\
        & O96     & 13.370 & $-0.072$ & $-0.94$ \\
        & R78     & 12.59  & $-0.09$  & $-0.95$ \\
        & V80     & 13.38  & $-0.07$  & $-0.95$ \\
        & Adopted & $13.375 \pm 0.005$ & $-0.071 \pm 0.001$ & $-0.945 \pm 0.005$  \\
D243-5  & MCPS    & 13.249 & $-0.215$ & $-0.786613$ \\
        & O96     & 13.979 & $-0.296$ & $-0.857$ \\
        & V80     & 13.54  & $-0.16$  & $-0.95$ \\
        & Adopted & $13.76 \pm 0.22$ & $-0.23 \pm 0.07$ & $-0.90 \pm 0.05$ \\
\enddata

\tablecomments{
Star identifiers follow the nomenclature of \citet{Oey1996UBV}. O96, R78, and V80 denote the photometry of \citet{Oey1996UBV}, \citet{Rousseau1978}, and \citet{vandenBergh1980}, respectively.
The uncertainties quoted for the adopted values represent
the spread among the literature measurements rather than
formal photometric errors.
}

\end{deluxetable*}

The extinction correction adopted throughout this work is based on the average value derived for hot stars in LH83 by \citet{Zaritsky2004}, corresponding to $A_V = 0.35$. 
This value can be compared with the extinction determined from the nebular
Balmer lines sampled across much of DEM~L243 \citep{Oey2000}. While the nebular
extinction spans a range of $A_V \sim 0$--0.7, typical values outside the 
immediate vicinity of the SNR are $A_V \sim 0.17$. 
To illustrate the potential impact of this lower extinction, the CMDs in Figures~\ref{fig:LH83_mcps_bv_summary} and \ref{fig:LH83_mcps_ub_corr_summary} 
include a reddening vector showing the displacement expected if $A_V = 0.17$ 
is adopted instead of $A_V = 0.35$.

In the extinction-corrected $V$ vs.\ $B-V$ CMD shown in Figure~\ref{fig:LH83_mcps_bv_summary},
most of the stars with spectral classifications are located near the evolutionary tracks; 
however, several objects, for example D243-2 and D243-5, appear bluer than the 
main sequence.
Adopting the lower extinction value of $A_V \sim 0.17$
would move several of these stars closer to the evolutionary tracks, 
as indicated by the reddening vector. 
However, the large uncertainties and the broad distribution of stars in this 
CMD make it difficult to derive robust masses from $B-V$ alone. 

In the extinction-corrected $B$ vs.\ $U-B$ CMD shown in Figure~\ref{fig:LH83_mcps_ub_corr_summary}, most of the stars classified by \citet{Oey1996spec} as O-type or early B-type stars are recovered in the region bounded by the 15--40~$M_\odot$ evolutionary tracks. 
The CMD-based masses are generally lower than the spectroscopy-based evolutionary-track estimates reported by \citet{Oey1996spec}, particularly for D243-2 and D243-3. 
A comparison between the CMD-based estimates and the spectroscopy-based evolutionary-track estimates is summarized in Table~\ref{tab:oey_phot_cmd_mass}.
As quantified in Section~3.1, the CMD-based masses are lower
than the HR-diagram masses inferred using the same MIST tracks
by $8.0~M_\odot$ on average, with a scatter of $7.0~M_\odot$
and an rms difference of $10.3~M_\odot$ for seven stars
(Figure~\ref{fig:cmd_vs_hr_mass}).
Nevertheless, the overall distribution of the spectroscopically identified stars is broadly consistent with the expected location of the massive stellar population in LH83.

\begin{figure}[htbp]
  \centering
  \includegraphics[width=0.45\textwidth]{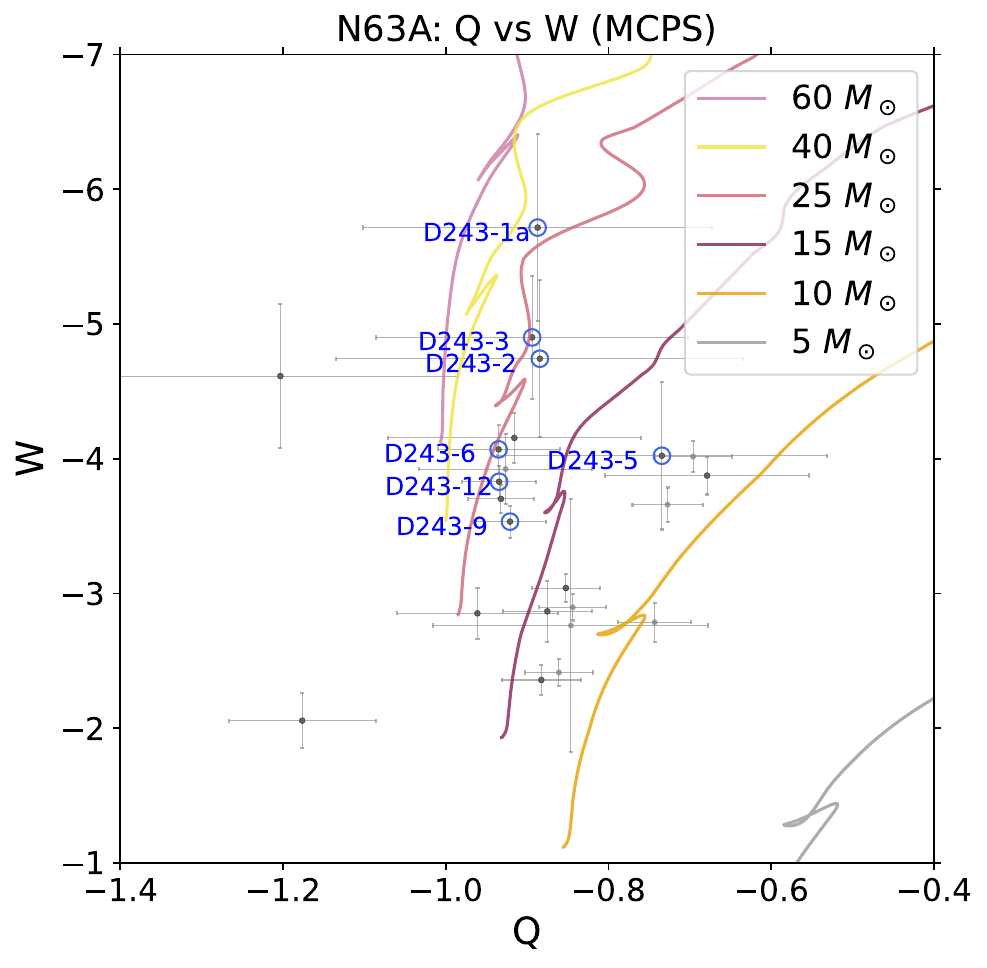}
    \caption{
    Reddening-free color--magnitude diagram of stars in the N63A region constructed from MCPS photometry ($W$ vs.\ $Q$).
    MIST evolutionary tracks are overplotted to estimate stellar masses while minimizing the effects of extinction.
    }
  \label{fig:LH83_mcps_qw}
\end{figure}

As a complementary diagnostic of the sensitivity to extinction, we also constructed a $W$ versus $Q$ diagram using the parameters described in Section~\ref{subsec:CMD_reddening_free}.
The resulting diagram is shown in Figure~\ref{fig:LH83_mcps_qw}.
Because the reddening-free transformation is most reliable for hot stars, only sources with $U < 14.5$ were included in this analysis.
The reddening-free CMD places most of the spectroscopically identified massive stars near the $15-40~M_\odot$ evolutionary tracks.

However, the reddening-free transformation substantially increases the propagated photometric uncertainties, resulting in larger error bars than in the conventional CMDs. 
For several stars, the uncertainties span multiple evolutionary tracks, limiting the precision of individual mass estimates. 

One notable exception is D243-5, which is placed at a lower mass than expected from its spectroscopic classification. 
This large discrepancy might be a result of the large photometric uncertainties.

A comparison of the three CMD diagnostics highlights the strengths and limitations of each method. 
The $V$ vs.\ $B-V$ CMD is strongly affected by both extinction uncertainties and photometric errors, leading to a broad stellar sequence and less robust mass estimates. 
In contrast, the $B$ vs.\ $U-B$ CMD provides a tighter distribution and recovers the majority of the spectroscopically identified massive stars within the expected mass range. 
The reddening-free CMD yields results broadly consistent with those obtained from the $B$ vs.\ $U-B$ diagram, as expected given the relatively low extinction toward LH83.
However, the propagated uncertainties in the reddening-free parameters are substantially larger than those in the conventional CMDs. 
We therefore adopt the $B$ vs.\ $U-B$ CMD as the primary basis for identifying massive-star candidates and estimating their masses in the following discussion.

\subsection{CMD from Gaia photometry}
\label{subsec:CMD_Gaia}

\begin{figure}[htbp]
  \centering
  \includegraphics[width=0.5\textwidth]{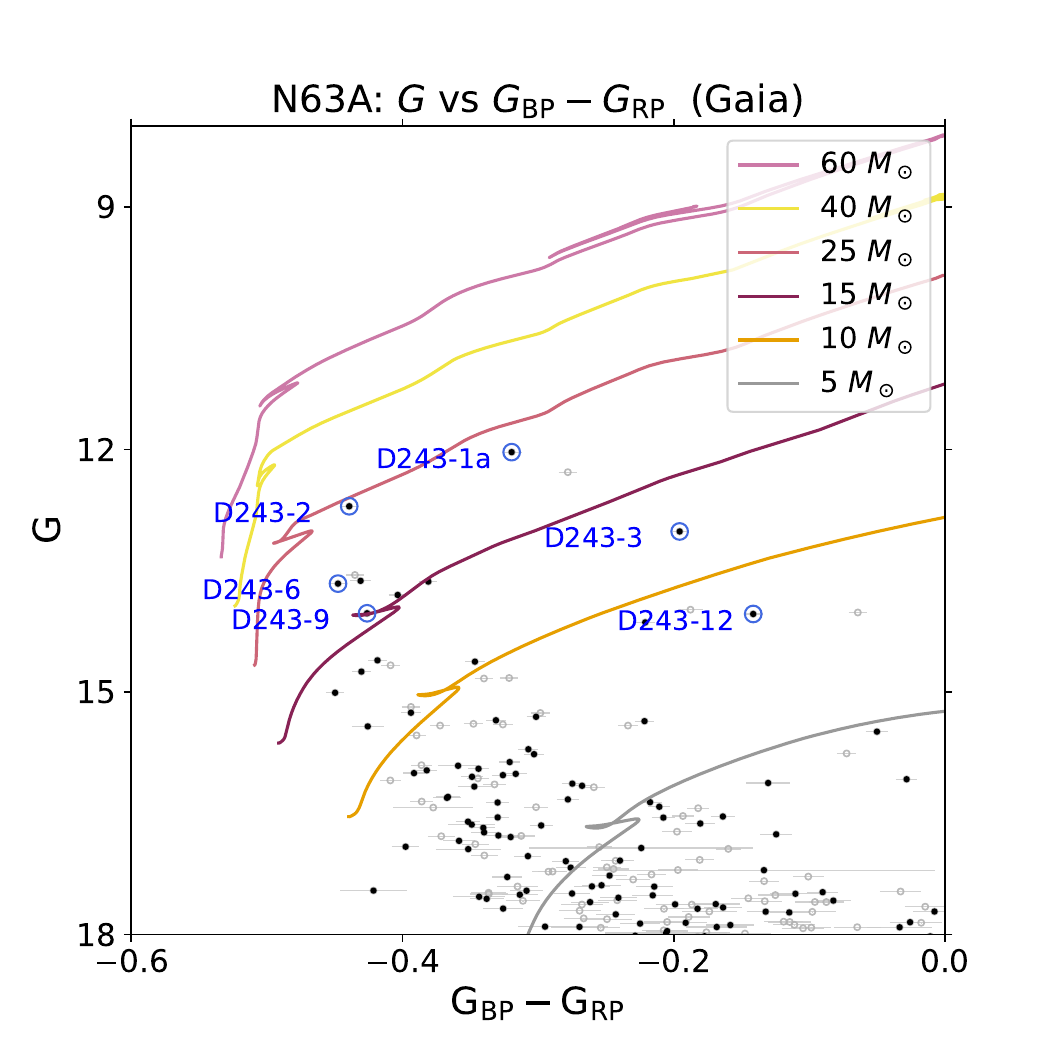}
  \vspace{1em}
  \includegraphics[width=0.48\textwidth]{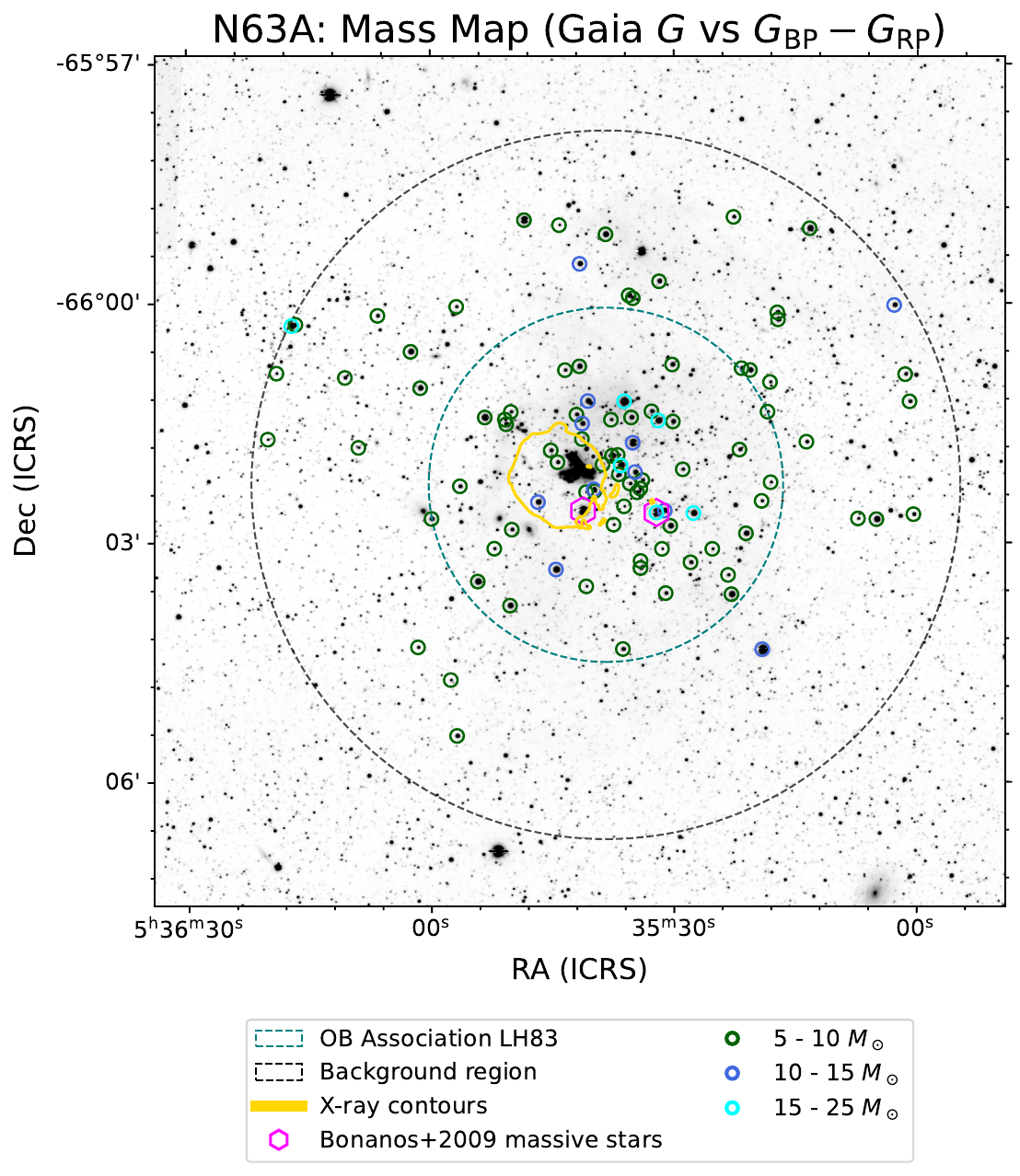}
  \caption{
    LH83: (Top) Color--magnitude diagram ($G$ vs.\ $G_\mathrm{BP}-G_\mathrm{RP}$) from Gaia EDR3.
    Stars within the OB association are shown in black, while those outside are shown in gray.
    (Bottom) Massive star candidates from Gaia EDR3 overlaid on the continuum image.
  }
  \label{fig:LH83_gaia}
\end{figure}

The Gaia EDR3 CMD is shown in Figure~\ref{fig:LH83_gaia}.
Gaia has a clear advantage for very bright stars. In particular, it recovers stars that are saturated or missing in MCPS. This makes Gaia an important complement to the MCPS dataset.

At the same time, the Gaia CMD does not separate the highest-mass stars as clearly as the MCPS $B$ vs.\ $U-B$ diagram. Several stars that are spectroscopically known to be massive are placed at lower inferred masses in the Gaia CMD.
To examine the effect of extinction, we also tested $A_V=0.17$, 0.35, 0.50, and 0.65 mag. The positions of the reference stars in the Gaia CMD change with the adopted extinction, affecting their inferred mass ranges, but the Gaia CMD does not provide robust individual mass estimates over the tested range.
The limited temperature sensitivity of the broad Gaia passbands, together with uncertainties in extinction and photometric measurements, may contribute to these discrepancies. 
We also examined the Gaia EDR3 photometric quality indicators for the spectroscopic reference stars.
Although the formal BP and RP photometric signal-to-noise ratios are high, all BP and RP transits of D243-6 are flagged as blended, indicating that its Gaia color may be affected by source confusion.
Source crowding and counterpart identification also require caution, particularly for D243-5, which has multiple Gaia sources within $2^{\prime\prime}$ and lacks usable $G_{\rm BP}$ and $G_{\rm RP}$ photometry.
Therefore, we use Gaia primarily to identify and recover bright stars that are missing or saturated in MCPS, while the MCPS $B$ vs.\ $U-B$ CMD remains our primary diagnostic for estimating stellar masses.

\section{Discussion} \label{sec:discussion}

\subsection{Age Estimate of OB Association LH83}

\begin{figure*}[htbp]
  \centering
  \includegraphics[width=0.48\textwidth]{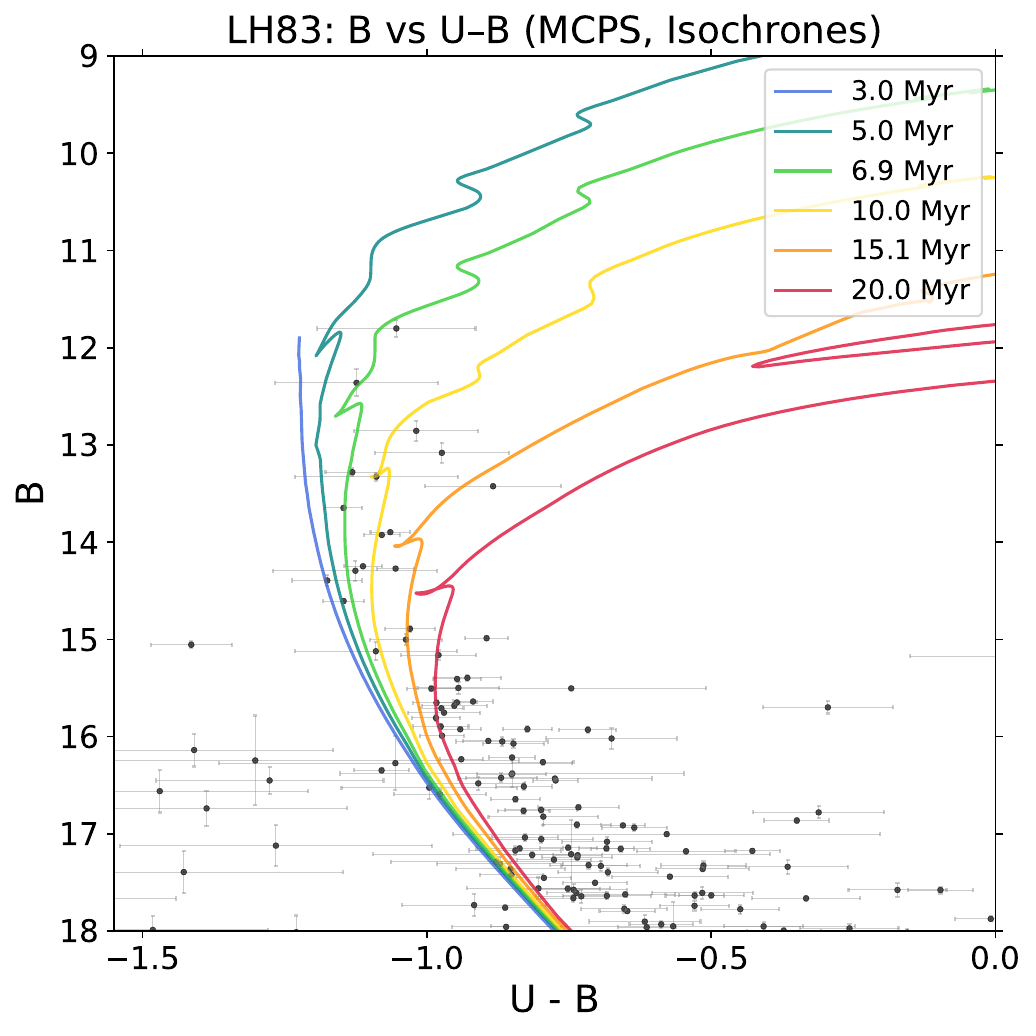}
  \hfill
  \includegraphics[width=0.48\textwidth]{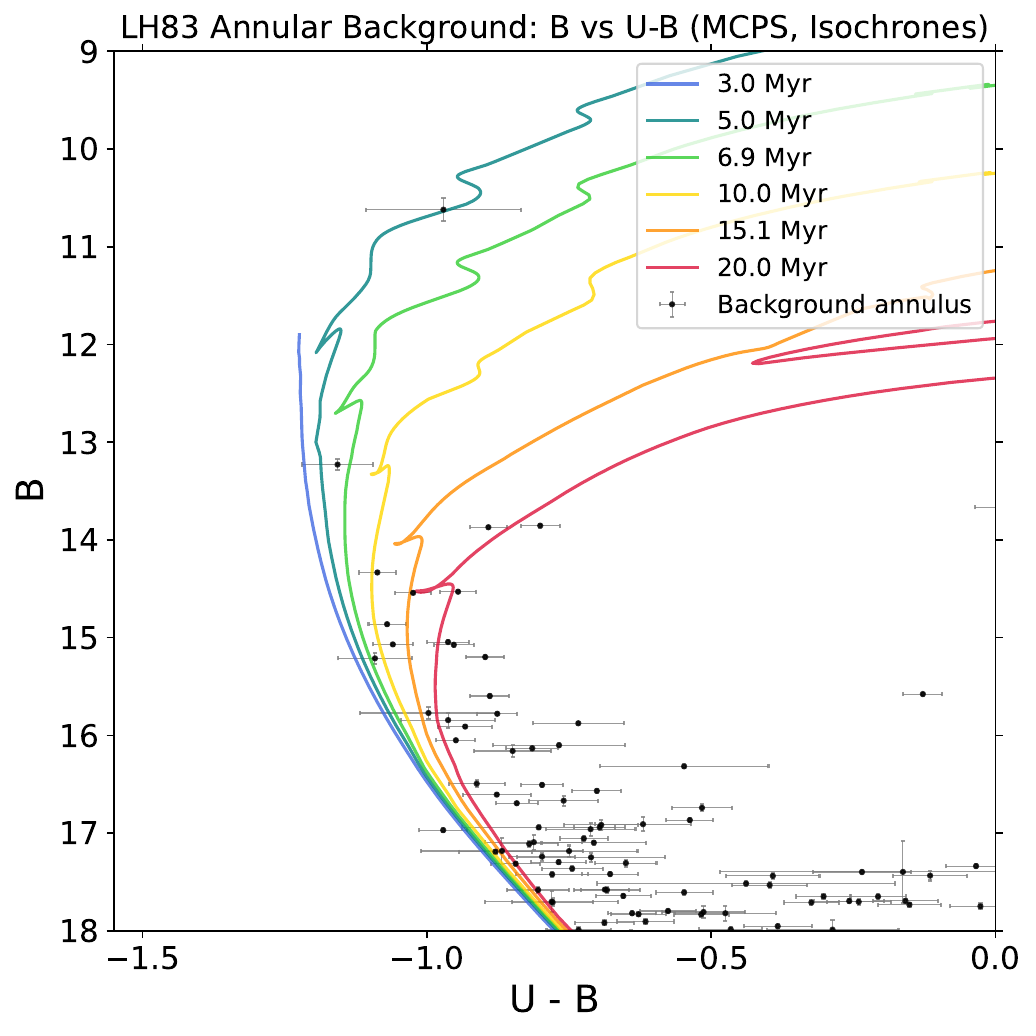}
  \caption{
  Comparison of MCPS $B$ vs.\ $U-B$ CMDs with MIST isochrones for the LH83 region and its surrounding background annulus. Left: CMD of stars located within the adopted LH83 boundary. Right: CMD of stars located only in the surrounding background annulus. In both panels, black points show the selected MCPS sources, and MIST isochrones with ages of 3, 5, 6.9, 10, 15.1, and 20 Myr are overplotted for comparison.
  }
  \label{fig:LH83_mcps_iso}
\end{figure*}

Because LH83 is superposed on a large-scale field population, both the LH83 region and the surrounding annular background region, defined in Section~\ref{sec:analysis}, need to be examined and compared.
Figure~\ref{fig:LH83_mcps_iso} shows the MCPS $B$ vs.\ $U-B$ CMDs overplotted with the MIST isochrones.
The stars in the LH83 and background regions are shown in the left and right panels, respectively.

Two prominent differences can be seen between the LH83 region and the background region.
First, in the $B=15$--18 range, the CMD of the LH83 region shows several stars with $U-B$ colors bluer than the main sequence, while the background region does not show such stars.
UV-bright stars may be young stars that are undergoing accretion; however, those
in LH83 are not identified as young stellar objects by \citet{Caulet2008} or located in
star-forming molecular clouds \citep{Sano2019}.  On the contrary, the 
UV-bright stars in LH83 are all found within the bright cloverleaf-shaped nebulosity. The close 
association between these UV-bright stars and bright nebulosity in N63A is indicative that the 
$U$ band photometry is significantly contaminated by the higher-order Balmer lines and [O~II] line 
emission from the nebula and that some of these sources may be nebular knots instead of real stars.
As these UV-bright sources are not massive stars, they will not be discussed further.

The second difference is seen in the $B<15$ range, where massive stars with initial masses 
greater than $\sim 10~M_\odot$ are located.
The background region shows that the great majority of massive field stars are bracketed by 
the 10 and 20 Myr isochrones in Figure~\ref{fig:LH83_mcps_iso}.
In contrast, the LH83 region shows stars extending from the 7 Myr to 15 Myr 
isochrones.  Considering that the background region has three times the surface area as the
LH83 region, we expect only $\sim$1 background massive star within the 7--15 Myr range in the LH83 region.
It may be concluded that most of the massive stars observed in the LH83 OB association were formed about 7--15 Myr ago.

Several factors can contribute to this apparent age spread in LH83, such as 
photometric errors, differential reddening, stellar binarity, and model uncertainties.  
It should also be noted that the formation of the OB association LH83 may not be instantaneous 
because of its large spatial extent, about 30 pc in radius.
If star formation started at one point and then propagated outward 
at the sound speed of a $10^4$~K ionized gas, i.e., $10~{\rm km~s^{-1}}$, it would extend to 
a distance of 30 pc in 3 Myr.  In an extreme example, the OB association LH15 extends over 400 pc, from the center to the rim of the supergiant shell LMC~1, and an age gradient in the stellar population is clearly seen \citep{Li2025}.
Thus, an age spread in an OB association is most likely real, and the age spread will inevitably add more uncertainty to the estimate of a supernova progenitor's mass.

\subsection{The Stellar Mass of N63A's SN Progenitor}

If a single burst of star formation follows an initial mass function, such as the Salpeter IMF \citep{Salpeter1955}, the observed present-day mass function can be used to extrapolate to higher stellar masses and estimate how many massive stars have evolved and vanished in supernovae or black holes.
If the star formation occurred over an extended period of time, the present-day mass function becomes more complicated.
The highest-mass stars from the oldest population may have vanished, while the lowest-mass stars from the youngest population may not yet have reached the main sequence.
To analyze the present-day mass function, it is essential to consider the stellar lifetimes alongside the population ages.

A useful check on the present-day mass function is the relative number of stars in adjacent mass intervals.
The CMD in Figure~\ref{fig:LH83_mcps_ub_corr_summary} shows 6 and 11 stars in the 10--15 and 15--25~$M_\odot$ ranges, respectively.
For a Salpeter IMF, $\xi(M) \propto M^{-2.35}$, lower-mass stars should be more numerous than higher-mass stars; the expected number ratio of 15--25 to 10--15~$M_\odot$ stars is therefore only $\sim 0.7$.
However, the observed ratio is 1.8.
If the number of stars in the background region is prorated and subtracted from these two mass ranges, the ratio becomes 2.7.
While these star counts may be subject to small-number statistics, it is possible that some of the most massive stars are so young that their less-massive siblings have not yet reached the main sequence.
Such an age effect could explain why the observed ratio is higher than the Salpeter expectation, reaching 2.7 after background subtraction.

In this case, their more-massive siblings, if they exist, should still be on the main sequence, and their absence does not mean that they have gone supernovae.
It is also questionable whether the Salpeter IMF is meaningful under this circumstance.
We therefore deem that the present-day mass function of LH83 cannot be used to extrapolate to higher stellar masses to estimate the number of massive stars that have exploded.
Nevertheless, we may find some clue to the SN progenitor mass based on the mass and evolutionary stage of the two most massive stars, D243-1a and D243-2.
Both of these stars have evolved off the main sequence and are located close to the 6.9 Myr isochrone.
Their masses are greater than $25~M_\odot$.
According to the MIST models we adopted for the LMC massive stars, the lifetimes (from
birth to the end) are 7.2 and 8.4 Myr for 30 and 25 $M_\odot$ stars, respectively.
It is conceivable that a more massive sibling of theirs, say, a $30~M_\odot$ star, was the SN progenitor of N63A.

The lack of an H$\alpha$ shell structure encompassing the X-ray emission from SNR N63A suggests that the SN exploded in the central cavity of a bubble blown by the fast stellar wind of the SN progenitor.
The wind-blown bubble itself may exhibit an expanding shell structure kinematically, but the expansion velocity is only up to twice the isothermal sound velocity and does not compress the ambient medium strongly enough to produce a morphologically identifiable shell \citep{Naze2001}.
Cavity explosions are common for core-collapse SNe. 
In the case of a cavity explosion, the SNR shocks expand quickly to the inner wall of the cavity, thus the X-ray size is a good approximation of the bubble size,
while the H~II region represents the ambient medium.

An empirical relationship between the cavity size and the progenitor mass has been reported by \citet{Chen2013}:
$P_5^{1/3} R_b = [\alpha (M/M_\odot) - \beta]~{\rm pc}$.
Here, $P_5$ is the pressure $nT$ normalized by $10^5~{\rm cm^{-3}~K}$, where $n$ is the particle number density in units of ${\rm cm^{-3}}$ and $T$ is the gas temperature in K.
The quantity $R_b$ is the bubble radius in pc, and $M$ is the stellar mass.
The constants are $\alpha = 1.22 \pm 0.05$ and $\beta = 9.16 \pm 1.77$.
If the ambient ISM density and the bubble size are known, the SN progenitor mass can be estimated from this relationship.

The diameter of N63A SNR is roughly 20 pc, and this is adopted as the bubble diameter.
The corresponding bubble radius is therefore $R_b = 10~{\rm pc}$.
The ambient ISM density can be diagnosed using the [S~II] $\lambda6716/\lambda6731$ line ratios from the spectroscopic observations of multiple slit positions by \citet{Oey2000}.
This [S~II] doublet ratio is near 1.4 throughout the H~II region except in the SNR itself, especially in the cloverleaf-shaped N63A region where molecular clouds are detected \citep{Sano2019}.
This indicates that the electron density, $n_e$, is $\sim 50~{\rm cm^{-3}}$ in N63 in general, but rises to $\sim 500~{\rm cm^{-3}}$ in the dense N63A region.
We adopt the general $n_e$ value, corresponding to a particle density of $n \sim 2n_e$ when protons are included.
We also adopt $T \sim 10^4~{\rm K}$ for the above bubble-size--progenitor-mass relationship.
This gives an expected progenitor mass of $\sim 25~M_\odot$.

The mass of N63A's SN progenitor has been independently estimated from analyses of X-ray spectra.  \citet{Karagoz2023} analyzed an archival 43 ks Chandra 
observation, modeled the spectra of SN ejecta and the ISM, and suggested that the O/Ne, 
O/Mg, and Ne/Mg ratios are consistent with a $\sim$40 $M_\odot$ SN progenitor.  In the meantime,
\citet{Chen2026} analyzed the same Chandra observation in conjunction with a 9.7 ks XMM-Newton
observation, and suggested an SN explosion energy of $4\times10^{51}$ ergs and an SN progenitor 
mass of $\sim$20 $M_\odot$.  These SN progenitor mass and explosion energy are both about half 
the values concluded by \citet{Karagoz2023}.  The large discrepancies between these two X-ray
analyses adequately illustrate that X-ray observations have not reached the desired spectral 
resolution for accurate line flux measurements or ionization equilibrium assessment.  While 
X-ray spectra analyses of SN ejecta can differentiate between core-collapse and Type Ia SN
progenitors, they lack the spectral resolution for accurate quantitative abundance determination 
to assess the SN progenitor mass.

\section{Conclusion} \label{sec:conclusion}

We have examined the stellar population around N63A using Gaia EDR3, MCPS, and published spectroscopic data for the OB association LH83.

Among the tested CMDs, the MCPS $B$ vs.\ $U-B$ diagram provides the most useful constraint on the massive-star population.
It provides a relatively tight stellar sequence and places most of the spectroscopically classified massive stars in the expected high-mass region of the CMD.
Nevertheless, the CMD-based masses are lower than the HR-diagram-based MIST masses by $8.0~M_\odot$ on average, with a scatter of $7.0~M_\odot$, indicating substantial uncertainties in individual stellar mass estimates.
In contrast, the $V$ vs.\ $B-V$ CMD is less reliable for identifying high-mass stars in this region because $B-V$ is only weakly sensitive to temperature in the hot-star regime and is more strongly affected by extinction and photometric uncertainties.
Gaia EDR3 is valuable for recovering very bright sources that are saturated or missing in MCPS, but its broad passbands provide weaker separation among the hottest stars and are less useful for precise mass estimation.
The reddening-free $W$--$Q$ diagram provides a useful check on the possible effects of extinction, but it does not separate the known massive stars more clearly than the MCPS $B$ vs.\ $U-B$ CMD.

The isochrone comparison indicates that most of the massive stars observed in LH83 were formed about 7--15 Myr ago.
Several factors can contribute to the apparent age spread,
such as photometric errors, differential reddening, unresolved binaries, and uncertainties in stellar
models; nevertheless, the 30 pc radius of the LH83 OB association requires the star formation to 
propagate at sound speed over 30 pc distance and an intrinsic age spread is inevitable.
The age spread makes it impossible to use the present-day mass function of LH83 alone to assess the
SN progenitor mass of N63A.  Additional constraints or clues are needed.

We considered (1) the CMD-based masses of D243-1a ($\sim$28 $M_\odot$) and D243-2 ($\sim$30 $M_\odot$), (2) the progenitor mass of approximately 25~$M_\odot$ estimated from the empirical bubble-size--progenitor-mass relation, and (3) the stellar lifetimes of 7.2 and 8.4 Myr predicted by the MIST models for stars of 30 and 25~$M_\odot$, respectively.
Taken together, these considerations suggest a SN progenitor mass of order 25--30~$M_\odot$.
This estimate represents an approximate mass scale rather than a formal confidence interval.
This mass scale is roughly consistent with the earlier estimate of a progenitor mass of at least 30~$M_\odot$ by \citet{vandenBergh1980}, and lies between the two progenitor mass estimates based on X-ray analyses of N63A, $\sim$40 $M_\odot$ \citep{Karagoz2023} and $\sim$20 $M_\odot$ \citep{Chen2026}.

Our analysis demonstrates the potential of the MCPS $B$ vs.\ $U-B$ CMD for identifying massive-star populations and constraining SN progenitor masses
in OB associations.
Future improvements in $U$-band photometry, together with more reliable extinction estimates for individual stars, will help reduce uncertainties in stellar mass estimates and further strengthen this approach for studies of other SNRs in the LMC.

%% Also note that the akcnowlodgment environment does not support long amounts of text. If you have a lot of people and institutions to acknowledge, do not use this command. Instead, create a new \section{Acknowledgments}.
\begin{acknowledgments}
This work has made use of data from the European Space Agency (ESA) mission {\it Gaia} (\url{https://www.cosmos.esa.int/gaia}), processed by the {\it Gaia} Data Processing and Analysis Consortium (DPAC, \url{https://www.cosmos.esa.int/web/gaia/dpac/consortium}). Funding for the DPAC has been provided by national institutions, in particular the institutions participating in the {\it Gaia} Multilateral Agreement.
%Funding for the DPAC has been provided by national institutions, in particular the institutions participating in the Gaia Multilateral Agreement.
This project is supported by the National Science and Technology Council grant NSTC 114-2811-M-110-037-, 114-2112-M-110-004-, 115-2112-M-110-001-, and 115-2811-M-110-015-.
\end{acknowledgments}


\vspace{5mm}
\facilities{CXO, Gaia, CTIO:4m, Swope}


\software{
Astropy \citep{2013A&A...558A..33A, Astropy2018, Astropy2022},
SciPy \citep{Virtanen2020},
NumPy \citep{Harris2020}
}



%\appendix


\bibliography{sample631}{}
\bibliographystyle{aasjournal}



\end{document}